\documentclass[preprint,superscriptaddress,nofootinbib]{revtex4}
  
  \usepackage{graphicx}
  \usepackage{rotating}
  \usepackage{multirow}
  \begin{document}
  \title{$J/{\psi}$ ${\to}$ $D_{s,d}{\pi}$, $D_{s,d}K$ decays with perturbative QCD approach}
  \author{Junfeng Sun}
  \affiliation{Institute of Particle and Nuclear Physics,
              Henan Normal University, Xinxiang 453007, China}
  \author{Yueling Yang}
  \affiliation{Institute of Particle and Nuclear Physics,
              Henan Normal University, Xinxiang 453007, China}
  \author{Jie Gao}
  \affiliation{Institute of Particle and Nuclear Physics,
              Henan Normal University, Xinxiang 453007, China}
  \author{Qin Chang}
  \affiliation{Institute of Particle and Nuclear Physics,
              Henan Normal University, Xinxiang 453007, China}
  \author{Jinshu Huang}
  \affiliation{College of Physics and Electronic Engineering,
              Nanyang Normal University, Nanyang 473061, China}
  \author{Gongru Lu}
  \affiliation{Institute of Particle and Nuclear Physics,
              Henan Normal University, Xinxiang 453007, China}

  %%%%%%%%%%%%%%%%%%%%%%%%%%%%%%%%%%%%%%%%%%%%%%%%%%%%%%%%%%%
  \begin{abstract}
  Besides the conventional strong and electromagnetic decay modes,
  the $J/{\psi}$ particle can also decay via the weak interaction
  in the standard model. In this paper, nonleptonic $J/{\psi}$
  ${\to}$ $D_{s,d}{\pi}$, $D_{s,d}K$ weak decays, corresponding
  to the externally emitted virtual $W$ boson process, are investigated
  with the perturbative QCD approach.
  It is found that branching ratio for the Cabibbo-favored
  $J/{\psi}$ ${\to}$ $D_{s}{\pi}$ decay can reach up to
  ${\cal O}(10^{-10})$, which might be potentially measurable
  at the future high-luminosity experiments.
  \end{abstract}
  \maketitle

  %%%%%%%%%%%%%%%%%%%%%%%%%%%%%%%%%%%%%%%%%%%%%%%%%%%%%%%%%%%
  \section{Introduction}
  \label{sec01}
  The discovery of the $J/{\psi}$ particle in 1974 at BNL in $p$-Be collisions
  \cite{bnl} and at SLAC in $e^{+}e^{-}$ collisions \cite{slac} provides
  evidence of the existence of the charm quark, and verifies that the quarks
  are physical elementary particles rather than purely mathematical
  entities \cite{gell}.
  The $J/{\psi}$ meson consists of the charm quark and antiquark
  pair $c\bar{c}$, so it carries some given quantum numbers, such
  as spin, isospin, parity, and charge conjugation, i.e.,
  $I^{G}J^{PC}$ $=$ $0^{-}1^{--}$ \cite{pdg}.
  The mass of $J/{\psi}$ meson is about three times the proton mass,
  but the width of $J/{\psi}$ meson is extremely narrow,
  only about 30 ppm\footnotemark[1] of its mass.
  One of the major reasons for the characteristic width is that $J/{\psi}$ ${\to}$
  $D\overline{D}$ is forbidden inasmuch as the $J/{\psi}$ meson lies below
  the kinematic $D\bar{D}$ threshold, and it is required by the
  $C$-parity conservation and the sacrosanct spin-statistics theorem
  that the $J/{\psi}$ meson must strongly decay into light hadrons
  via the $c\bar{c}$ annihilation into three gluons, which is of higher
  order in the quark-gluon coupling ${\alpha}_{s}$ and is therefore
  suppressed by the phenomenological Okubo-Zweig-Iizuka (OZI) rules \cite{o,z,i}.
  \footnotetext[1]{ppm means percent per million, i.e., $10^{-6}$.}

  The $J/{\psi}$ decay modes are usually partitioned into four categories:
  hadronic decay $J/{\psi}$ ${\to}$ $ggg$ with branching ratio
  ${\sim}$ $(64.1{\pm}1.0)\%$ \cite{pdg}, electromagnetic decay
  $J/{\psi}$ ${\to}$ ${\gamma}^{\ast}$ with branching ratio ${\sim}$
  $(2+R){\cal B}r_{ee}$, radiative decay $J/{\psi}$ ${\to}$
  ${\gamma}gg$ with branching ratio ${\sim}$ $(8.8{\pm}1.1)\%$ \cite{pdg},
  and magnetic dipole transition decay $J/{\psi}$ ${\to}$ ${\gamma}{\eta}_{c}$
  with branching ratio ${\sim}$ $(1.7{\pm}0.4)\%$ \cite{pdg},
  where the ratio of the production cross section $R$ $=$
  ${\sigma}(e^{+}e^{-}{\to}X)/{\sigma}(e^{+}e^{-}{\to}{\mu}^{+}{\mu})$
  and ${\cal B}r_{ee}$ is the branching ratio for the pure leptonic $J/{\psi}$
  ${\to}$ $e^{+}e^{-}$ decay. Because of OZI rule violation,
  the electromagnetic $J/{\psi}$ decay can
  compete favorably with hadronic $J/{\psi}$ decay. The properties
  of gluons and the quark-gluon coupling can be collected in
  hadronic and radiative $J/{\psi}$ decay. In addition, the radiative
  $J/{\psi}$ decay offers an ideal plaza to search for possible
  glueballs. Besides, the $J/{\psi}$ decay via the weak interaction
  is permissible within the standard model.
  In this paper, we will investigate the charm-changing nonleptonic
  $J/{\psi}$ ${\to}$ $D_{s,d}{\pi}$, $D_{s,d}K$ weak decays with the
  perturbative QCD (pQCD) approach \cite{pqcd1,pqcd2,pqcd3}.
  Our motivation is listed as follows.

  Experimentally,
  (1) thanks to the tremendous impetus from BES, CLEO-c, B-factories,
  LHCb, and so on, the $J/{\psi}$ particle attracts much persistent
  attention of experimentalists and theorists.
  A large amount of $J/{\psi}$ data samples have been accumulated.
  It is promisingly expected to produce about $10^{10}$ $J/{\psi}$
  samples at BESIII per year with the designed luminosity \cite{cpc36},
  and over $10^{10}$ prompt $J/{\psi}$ samples at LHCb per $fb^{-1}$
  data \cite{epjc71}.
  It is not utopian to carefully scrutinize the $J/{\psi}$ weak decays
  at the high-luminosity dedicated experiments in the future.
  (2) The production and ``flavor tag'' of single charged $D$ meson
  from $J/{\psi}$ decay will single the potential signal out from
  massive and intricate background.
  Recently, the $J/{\psi}$ ${\to}$ $D_{s}{\rho}$,
  $D_{u}K^{\ast}$ decays have been investigated at BESIII using
  $2.25{\times}10^{8}$ $J/{\psi}$ data samples \cite{prd89.071101},
  although no evidence is found due to tiny accident probabilities
  and insufficient available data samples. It is hard but interesting
  to hunt for $J/{\psi}$ weak decay experimentally.
  A deviant production rate of single $D$ meson from $J/{\psi}$
  decay would be a hint of new physics.

  Theoretically, the $J/{\psi}$ ${\to}$ $D_{q}P$ decay is
  induced by $c$ ${\to}$ $q$ $+$ $W^{+}$ transition, where
  $q$ $=$ $s$ and $d$, and the virtual $W^{+}$ boson materializes
  into a pair of quarks which then grows into a pseudoscalar
  meson $P$ $=$ ${\pi}$ and $K$. As it is well known,
  nonleptonic $J/{\psi}$ weak decay must be with the participation
  of the strong interaction, and the $c$ quark mass is between
  nonperturbative and perturbative domain.
  Recently, many QCD-inspired methods have been
  developed, such as the pQCD approach \cite{pqcd1,pqcd2,pqcd3},
  the QCD factorization (QCDF) approach \cite{qcdf1,qcdf2,qcdf3},
  the soft and collinear effective theory \cite{scet1,scet2,scet3,scet4},
  and have been applied preferably to accommodate measurements on nonleptonic
  $B$ decays. Based on collinear approximation, the $J/{\psi}$
  ${\to}$ $DP$ decays have been studied with naive factorization
  \cite{plb252,ijmpa14,ahep2013} and the QCDF approach \cite{ijmpa30},
  where theoretical results differ mainly from hadronic input
  parameters. In this paper, the $J/{\psi}$ ${\to}$ $D_{s,d}P$
  decays will be studied with the pQCD approach based on $k_{T}$
  factorization. It is expected that with nonleptonic $J/{\psi}$
  weak decay, one can glean new insights into the factorization
  mechanism, nonfactorizable contributions, nonperturbative
  dynamics, final state interactions, and so on.

  This paper is organized as follows.
  In section \ref{sec02}, we present the theoretical framework
  and the amplitudes for the $J/{\psi}$ ${\to}$ $D_{s,d}P$ decay with
  the pQCD approach. Section \ref{sec03} is devoted to numerical
  results and discussion. The last section is our summary.

  %%%%%%%%%%%%%%%%%%%%%%%%%%%%%%%%%%%%%%%%%%%%%%%%%%%%%%%%%%%
  \section{theoretical framework}
  \label{sec02}
  %%%%%%%%%%%%%%%%%%%%%%%%%%
  \subsection{The effective Hamiltonian}
  \label{sec0201}
  Constructed by means of the operator product expansion and
  the renormalization group (RG) method, the effective Hamiltonian
  describing the $J/{\psi}$ ${\to}$ $D_{s,d}P$ weak decay could
  be written as a series of effective local operators $Q_{i}$
  multiplied by effective Wilson coefficients $C_{i}$ and
  have the following structure \cite{9512380}:
 %---------------------------------------------------------
   \begin{equation}
  {\cal H}_{\rm eff}\ =\ \frac{G_{F}}{\sqrt{2}}\,
   \sum\limits_{q_{1},q_{2}} V_{cq_{1}}V_{uq_{2}}^{\ast}\,
   \Big\{ C_{1}({\mu})\,Q_{1}({\mu})
         +C_{2}({\mu})\,Q_{2}({\mu}) \Big\}
   + {\rm h.c.}
   \label{hamilton},
   \end{equation}
 %---------------------------------------------------------
  where $G_{F}$ is the Fermi coupling constant and $q_{1,2}$ $=$ $d$, $s$.

  Using the Wolfenstein parameterization \cite{prl51}, there are
  some hierarchy relations among the Cabibbo-Kobayashi-Maskawa
  \cite{prl10,ptp49} (CKM) factors, i.e.,
 %---------------------------------------------------------
  \begin{eqnarray}
  V_{cs}V_{ud}^{\ast} &=&
   1-{\lambda}^{2}-\frac{1}{2}A^{2}{\lambda}^{4}
  +{\cal O}({\lambda}^{6})
  \label{eq:vcsvud}, \\
 %---------------------------------------------------------
  V_{cs}V_{us}^{\ast} &=&
   {\lambda}-\frac{1}{2}{\lambda}^{3}-\frac{1}{8}{\lambda}^{5}(1+4A^{2})
  +{\cal O}({\lambda}^{6})
  \label{eq:vcsvus}, \\
 %---------------------------------------------------------
  V_{cd}V_{ud}^{\ast} &=&
  -V_{cs}V_{us}^{\ast}-A^{2}{\lambda}^{5}({\rho}+i{\eta})
  +{\cal O}({\lambda}^{6})
  \label{eq:vcdvud}, \\
 %---------------------------------------------------------
  V_{cd}V_{us}^{\ast} &=&
  -{\lambda}^{2}+{\cal O}({\lambda}^{6})
  \label{eq:vcdvus},
  \end{eqnarray}
 %---------------------------------------------------------
  for $J/{\psi}$ ${\to}$ $D_{s}{\pi}$, $D_{s}K$, $D_{d}{\pi}$, $D_{d}K$
  decays, respectively,
  where the Wolfenstein parameter ${\lambda}$ $=$ ${\sin}{\theta}_{c}$
  ${\simeq}$ $0.2$ \cite{pdg} and ${\theta}_{c}$ is the Cabibbo angle.

  The auxiliary scale ${\mu}$ in Eq.(\ref{hamilton}) factorizes
  contributions into long- and short-distance dynamics.
  The Wilson coefficients $C_{1,2}(\mu)$ summarize the
  short-distance physical contributions above the scales
  of ${\mu}$. They are computable at the scale of
  the $W$ boson mass ${\mu}$ $=$ ${\cal O}(m_{W})$ with
  perturbation theory, and then evolved down to a
  characteristic scale for $c$ quark decay.
  %-----------------------------------------------------
  \begin{equation}
  \vec{C}({\mu}) = U_{4}({\mu},m_{b})U_{5}(m_{b},m_{W})\vec{C}(m_{W})
  \label{ci},
  \end{equation}
  %-----------------------------------------------------
  where $U_{f}({\mu}_{f},{\mu}_{i})$ is the RG evolution matrix
  transforming the Wilson coefficients from scale ${\mu}_{i}$
  to ${\mu}_{f}$. The explicit expression of $U_{f}({\mu}_{f},{\mu}_{i})$
  can be found in Ref.\cite{9512380}.
  The Wilson coefficients have properly been evaluated to
  the next-to-leading order.

  The penguin contributions are severely suppressed by the CKM
  factors $V_{cd}V_{ud}^{\ast}$ $+$ $V_{cs}V_{us}^{\ast}$ $=$
  $-V_{cb}V_{ub}^{\ast}$ ${\sim}$ ${\cal O}({\lambda}^{5})$,
  which are negligible in our calculation.
  Only the tree operators related to $W$ emission contributions
  are considered.
  The expressions of tree operators are
 %-----------------------------------------------------
    \begin{eqnarray}
    Q_{1} &=&
  [ \bar{q}_{1,{\alpha}}{\gamma}_{\mu}(1-{\gamma}_{5})c_{\alpha} ]
  [ \bar{u}_{\beta} {\gamma}^{\mu}(1-{\gamma}_{5})q_{2,{\beta}} ]
    \label{q1}, \\
 %-----------------------------------------------------
    Q_{2} &=&
  [ \bar{q}_{1,{\alpha}}{\gamma}_{\mu}(1-{\gamma}_{5})c_{\beta} ]
  [ \bar{u}_{\beta} {\gamma}^{\mu}(1-{\gamma}_{5})q_{2,{\alpha}} ]
    \label{q2},
    \end{eqnarray}
 %-----------------------------------------------------
  where ${\alpha}$ and ${\beta}$ are color indices and
  the sum over repeated indices is understood.
  The physical contributions below scales of ${\mu}$ are
  included in hadronic matrix elements (HME), where the local
  operators are sandwiched between initial and final hadron states.
  Generally, HME is the most complicated and intractable part,
  where the perturbative and nonperturbative effects entangle
  with each other. In addition, nonfactorizable corrections
  to HME should be taken into account decently so that the
  ${\mu}$ dependences of HME could cancel and/or milden those
  of Wilson coefficients.

  %%%%%%%%%%%%%%%%%%%%%%%%%%
  \subsection{Hadronic matrix elements}
  \label{sec0202}
  With the Lepage-Brodsky approach for exclusive processes \cite{prd22},
  HME could be expressed as the convolution of a hard scattering kernel
  with distribution amplitudes (DA) in parton momentum fractions, where
  DA reflecting the nonperturbative contributions is commonly assumed
  to be universal, which makes the structure simple. The hard part
  could be perturbatively computed in an expansion of strong
  coupling ${\alpha}_{s}$. Unfortunately, soft endpoint contributions
  do not admit self-consistent treatment with collinear factorization
  approximation \cite{qcdf1,qcdf2,qcdf3}.
  To settle the issue, in evaluation of potentially infrared
  contributions with the pQCD approach, the transverse momentum of
  quarks are kept explicitly and the Sudakov factors are
  introduced for each of mesonic DA \cite{pqcd1,pqcd2,pqcd3}.
  Finally, the decay amplitudes could be factorized into three
  parts \cite{pqcd2,pqcd3}: the hard effects enclosed by Wilson
  coefficients $C_{i}$, the heavy quark decay amplitudes
  ${\cal H}$, and process-independent wave functions ${\Phi}$,
  %-----------------------------------------------------
  \begin{equation}
  {\int} dk\,
  C_{i}(t)\,{\cal H}(t,k)\,{\Phi}(k)\,e^{-S}
  \label{hadronic},
  \end{equation}
  %-----------------------------------------------------
  where $t$ is a typical scale, $k$ is the momentum of valence
  quarks, and the Sudakov factor $e^{-S}$ is used to suppress the
  long-distance contributions and makes the hard scattering
  subprocess more perturbative.

  %%%%%%%%%%%%%%%%%%%%%%%%%%
  \subsection{Kinematic variables}
  \label{sec0203}
  In the rest frame of the $J/{\psi}$ meson, kinematic
  variables are defined as below:
  %------------------------------------
  \begin{equation}
  p_{J/{\psi}}\, =\, p_{1}\, =\, \frac{m_{1}}{\sqrt{2}}(1,1,0)
  \label{kine-p1},
  \end{equation}
  %------------------------------------
  \begin{equation}
  p_{D}\, =\, p_{2}\, =\, (p_{2}^{+},p_{2}^{-},0)
  \label{kine-p2},
  \end{equation}
  %------------------------------------
  \begin{equation}
  p_{P}\, =\, p_{3}\, =\, (p_{3}^{-},p_{3}^{+},0)
  \label{kine-p3},
  \end{equation}
  %------------------------------------
  \begin{equation}
  k_{i}\, =\, x_{i}\,p_{i}+(0,0,\vec{k}_{iT})
  \label{kine-ki},
  \end{equation}
  %------------------------------------
  \begin{equation}
  {\epsilon}_{\psi}^{\parallel}\, =\,
  {\epsilon}_{1}^{\parallel}\, =\,
   \frac{1}{\sqrt{2}}(-1,1,0)
  \label{kine-longe},
  \end{equation}
  %------------------------------------
  %------------------------------------
  \begin{equation}
  n_{+}=(1,0,0)
  \label{kine-null-plus},
  \end{equation}
  %------------------------------------
  \begin{equation}
  n_{-}=(0,1,0)
  \label{kine-null-minus},
  \end{equation}
  %------------------------------------
  \begin{equation}
  p_{i}^{\pm}\, =\, (E_{i}\,{\pm}\,p)/\sqrt{2}
  \label{kine-pipm},
  \end{equation}
  %------------------------------------
  \begin{equation}
  t\, =\, 2\,p_{1}{\cdot}p_{2}\, =\ 2\,m_{1}\,E_{2}
  \label{kine-t},
  \end{equation}
  %------------------------------------
  \begin{equation}
  u\, =\, 2\,p_{1}{\cdot}p_{3}\, =\ 2\,m_{1}\,E_{3}
  \label{kine-u},
  \end{equation}
  %------------------------------------
  \begin{equation}
  s\, =\, 2\,p_{2}{\cdot}p_{3}
  \label{kine-s},
  \end{equation}
  %------------------------------------
  \begin{equation}
  p = \frac{\sqrt{ [m_{1}^{2}-(m_{2}+m_{3})^{2}]\,[m_{1}^{2}-(m_{2}-m_{3})^{2}] }}{2\,m_{1}}
  \label{kine-pcm},
  \end{equation}
  %------------------------------------
  where the subscripts $i$ $=$ $1$, $2$, $3$ on variables $p_{i}$, $E_{i}$
  and $m_{i}$ correspond to $J/{\psi}$, $D$, $P$ mesons, respectively;
  $p_{i}$ is a four-dimensional momentum abiding by the on-shell
  condition $p_{i}^{2}$ $=$ $m_{i}^{2}$;
  $x_{i}$ and $k_{i}$ ($\vec{k}_{iT}$) denote the longitudinal momentum
  fraction and (transverse) momentum of a relatively light valence
  quark in mesons,
  respectively; ${\epsilon}_{1}^{\parallel}$ denotes the longitudinal
  polarization vector satisfying with the relations
  ${\epsilon}_{1}^{\parallel}{\cdot}{\epsilon}_{1}^{\parallel}$
  $=$ $-1$ and ${\epsilon}_{1}^{\parallel}{\cdot}p_{1}$ $=$ $0$;
  $n_{+}$ and $n_{-}$ are the plus and minus null vectors, respectively,
  complying with $n_{\pm}^{2}$ $=$ $0$ and $n_{+}{\cdot}n_{-}$ $=$ $1$;
  $t$, $u$, and $s$ are the Lorentz-invariant variables; and $p$ is the
  common momentum of the final states.
  The notation of momentum is displayed in Fig.\ref{feynman}(a).

  %%%%%%%%%%%%%%%%%%%%%%%%%%
  \subsection{Wave functions}
  \label{sec0204}
  Taking the convention of Refs. \cite{prd65,npb529}, the HME of the diquark
  operators squeezed between the vacuum and meson state
  is defined as below:
  %------------------------------------
  \begin{equation}
 {\langle}0{\vert}c_{i}(z)\bar{c}_{j}(0){\vert}
 {\psi}(p_{1},{\epsilon}_{1}^{\parallel}){\rangle} =
  \frac{f_{\psi}}{4}{\int}d^{4}k_{1}\,e^{-ik_{1}{\cdot}z}
  \Big\{ \!\!\not{\epsilon}_{1}^{\parallel} \Big[
   m_{1}\,{\Phi}_{\psi}^{v}(k_{1})
  -\!\!\not{p}_{1}\, {\Phi}_{\psi}^{t}(k_{1})
  \Big] \Big\}_{ji}
  \label{wave-cc},
  \end{equation}
  %------------------------------------
  \begin{equation}
 {\langle}D_{q}(p_{2}){\vert}\bar{q}_{i}(z)c_{j}(0){\vert}0{\rangle} =
  \frac{if_{D_{q}}}{4}{\int}d^{4}k_{2}\,e^{ik_{2}{\cdot}z}\,
  \Big\{ {\gamma}_{5}\Big[ \!\!\not{p}_{2}\,{\Phi}_{D}^{a}(k_{2})
  +m_{2}\,{\Phi}_{D}^{p}(k_{2})\Big] \Big\}_{ji}
  \label{wave-ds},
  \end{equation}
  %------------------------------------
  \begin{equation}
 {\langle}P(p_{3})
 {\vert}q_{i}(0)\bar{q}^{\prime}_{j}(z){\vert}0{\rangle} =
  \frac{if_{P}}{4}{\int}_{0}^{1}dk_{3}\,e^{ik_{3}{\cdot}z}
 \Big\{ {\gamma}_{5}\Big[ \!\!\not{p}_{3}\,{\Phi}_{P}^{a}(k_{3})
  +{\mu}_{P}\,{\Phi}_{P}^{p}(k_{3})
  +{\mu}_{P}\,(\!\not{n}_{-}\!\!\not{n}_{+}-1)\,{\Phi}_{P}^{t}(k_{3})
  \Big] \Big\}_{ji}
  \label{wave-pi},
  \end{equation}
  %------------------------------------
  where $f_{\psi}$, $f_{D_{q}}$, and $f_{P}$ are decay constants;
  wave functions ${\Phi}_{\psi}^{v}$ and ${\Phi}_{D,P}^{a}$
  are twist-2; wave functions ${\Phi}_{\psi}^{t}$ and
  ${\Phi}_{D,P}^{p,t}$ are twist-3. For the $J/{\psi}$ meson,
  the transverse polarization components contribute nothing
  to decay amplitudes in question.

  The decay constant $f_{\psi}$ can be obtained from the experimental
  branching ratios of the electromagnetic $J/{\psi}$ decay into charged
  lepton pairs through the formula
 %-----------------------------------------------------
   \begin{equation}
  {\cal B}r(J/{\psi}{\to}{\ell}^{+}{\ell}^{-}) =
   \frac{16{\pi}}{27}f_{\psi}^{2}
   \frac{{\alpha}_{\rm QED}^{2}}{m_{\psi}\,{\Gamma}_{\psi}}
   \sqrt{ 1-2\frac{m_{\ell}^{2}}{m_{\psi}^{2}} }
   \Big\{ 1+2\frac{m_{\ell}^{2}}{m_{\psi}^{2}} \Big\}
   \label{eq:fjpsi},
   \end{equation}
 %-----------------------------------------------------
  where ${\alpha}_{\rm QED}$ is the fine-structure
  constant, $m_{\ell}$ is the lepton mass and ${\ell}$
  $=$ $e$, ${\mu}$. Here, we will use the weighted average
  decay constant $f_{\psi}$ $=$ $395.1{\pm}5.0$ MeV
  (see Table \ref{tab:fjpsi}).

  %---------------------------------------------------------
   \begin{table}[h]
   \caption{Experimental branching ratios for leptonic
   $J/{\psi}$ decay and decay constant $f_{\psi}$, where
   ${\langle}f_{\psi}{\rangle}$ denotes the weighted average, and
   errors of decay constant arise from mass $m_{\psi}$,
   decay width ${\Gamma}_{\psi}$ and branching ratios.}
   \label{tab:fjpsi}
   \begin{ruledtabular}
   \begin{tabular}{cccc}
   decay mode & branching ratio & $f_{\psi}$ & ${\langle}f_{\psi}{\rangle}$ \\ \hline
    $J/{\psi}$ ${\to}$ $e^{+}e^{-}$
  & $(5.971{\pm}0.032)\%$
  & $395.4{\pm} 7.0 $ MeV &
    \multirow{2}{*}{$395.1{\pm} 5.0 $ MeV} \\
    $J/{\psi}$ ${\to}$ ${\mu}^{+}{\mu}^{-}$
  & $(5.961{\pm}0.033)\%$
  & $394.8{\pm} 7.1 $ MeV
   \end{tabular}
   \end{ruledtabular}
   \end{table}
  %---------------------------------------------------------

  For the emitted pseudoscalar $P$ meson, only the twist-2 wave functions
  are involved in the actual calculation (see Appendix \ref{blocks}).
  The twist-2 DA has the expansion \cite{npb529}:
  %------------------------------------
  \begin{equation}
 {\phi}_{P}^{a}(x) = 6\,x\,\bar{x}\, \Big\{ 1+
  \sum\limits_{i=1} a_{i}^{P}\,C_{i}^{3/2}(t) \Big\}
  \label{da-pi},
  \end{equation}
  %------------------------------------
  where $\bar{x}$ $=$ $1$ $-$ $x$ and $t$ $=$ $1$ $-$ $2x$;
  Gegenbauer moments $a_{i}^{P}$ corresponding to Gegenbauer polynomials
  $C_{i}^{3/2}(t)$ could be determined experimentally or with
  nonperturbative methods (such as QCD sum rules).
  It follows that $a_{i}^{P}$ $=$ $0$ for {\em odd} $i$ due to the $G$-parity
  invariance of DA for ${\pi}$ and ${\eta}^{(\prime)}$ mesons.
  Gegenbauer polynomials $C_{i}^{3/2}(t)$ have the expression
  %------------------------------------
  \begin{equation}
  C_{1}^{3/2}(t) = 3\,t,
  \quad
  C_{2}^{3/2}(t) = \frac{3}{2}\,(5\,t^{2}-1),
  \quad
  {\cdots}
  \label{polynomials}
  \end{equation}
  %------------------------------------

  Because of $m_{J/{\psi}}$ ${\simeq}$ $2m_{c}$ and $m_{D_{q}}$
  ${\simeq}$ $m_{c}$ $+$ $m_{q}$, it is commonly assumed that
  valence quarks in the charmonium $J/{\psi}$ and charmed
  mesons might be nearly nonrelativistic.
  Nonrelativistic quantum chromodynamics (NRQCD)
  \cite{prd46,prd51,rmp77} and the Schr\"{o}dinger
  equation can be used to describe their spectrum.
  The ground state eigenfunction of the time-independent
  Schr\"{o}dinger equation with an isotropic harmonic
  oscillator potential\footnotemark[2],
  \footnotetext[2]{A long time ago, many forms
  of phenomenological potential have been proposed to describe wave
  functions for heavy quarkonium states (such as $c\bar{c}$
  and $b\bar{b}$), for example, see Ref.\cite{prd52}.
  An isotropic harmonic oscillator is just a first approximation
  of potential for a stable system. Of course, this approximation
  is very rough. A more careful study of wave functions is always
  worthwhile but is beyond the scope of this paper.}
  corresponding to the quantum
  numbers $nL$ $=$ $1S$, has the form shown below in the
  momentum space,
  %-----------------------------------------------------
   \begin{equation}
  {\phi}_{1S}(\vec{k})\
  {\sim}\ e^{-\vec{k}^{2}/2{\omega}^{2}}
   \label{wave-k},
   \end{equation}
  %-----------------------------------------------------
  where parameter ${\omega}$ determines the average transverse
  momentum, ${\langle}1S{\vert}k^{2}_{T}{\vert}1S{\rangle}$
  $=$ ${\omega}^{2}$. According to the NRQCD power counting rules
  \cite{prd46}, the typical momentum is $k$ ${\sim}$ ${\omega}$
  ${\sim}$ $mv$ ${\sim}$ $m{\alpha}_{s}$, and the quark velocity
  $v$ is approximately equal to the effective QCD coupling strength
  ${\alpha}_{s}$. Employing the substitution transformation \cite{xiao},
  %-----------------------------------------------------
   \begin{equation}
   \vec{k}^{2}\ {\to}\ \frac{1}{4} \Big(
   \frac{\vec{k}_{T}^{2}+m_{q_{1}}^{2}}{x_{1}}
  +\frac{\vec{k}_{T}^{2}+m_{q_{2}}^{2}}{x_{2}} \Big)
   \label{wave-kt},
   \end{equation}
  %-----------------------------------------------------
  where $x_{i}$ fitting with $x_{1}$ $+$ $x_{2}$ $=$ $1$ is
  the longitudinal momentum fraction of valence quark with
  mass $m_{q_{i}}$, and then integrating out transverse momentum
  $k_{T}$ and combining with their asymptotic forms,
  one can obtain DA for $J/{\psi}$ and $D$ mesons,
  %-----------------------------------------------------
   \begin{equation}
  {\phi}_{\psi}^{v}(x) = A\, x\,\bar{x}\,
  {\exp}\Big\{ -\frac{m_{c}^{2}}{8\,{\omega}_{1}^{2}\,x\,\bar{x}} \Big\}
   \label{wave-ccv},
   \end{equation}
  %-----------------------------------------------------
   \begin{equation}
  {\phi}_{\psi}^{t}(x) = B\, t^{2}\,
  {\exp}\Big\{ -\frac{m_{c}^{2}}{8\,{\omega}_{1}^{2}\,x\,\bar{x}} \Big\}
   \label{wave-cct},
   \end{equation}
  %-----------------------------------------------------
   \begin{equation}
  {\phi}_{D}^{a}(x) = C\, x\,\bar{x}\, {\exp}\Big\{
  -\frac{\bar{x}\,m_{q}^{2}+x\,m_{c}^{2}}
        {8\,{\omega}_{2}^{2}\,x\,\bar{x}} \Big\}
   \label{wave-dqa},
   \end{equation}
  %-----------------------------------------------------
   \begin{equation}
  {\phi}_{D}^{p}(x) = D\, {\exp}\Big\{
  -\frac{\bar{x}\,m_{q}^{2}+x\,m_{c}^{2}}
        {8\,{\omega}_{2}^{2}\,x\,\bar{x}} \Big\}
   \label{wave-dqp},
   \end{equation}
  %-----------------------------------------------------
   where parameter ${\omega}_{i}$ $=$ $m_{i}{\alpha}_{s}(m_{i})$,
   and coefficients of $A$, $B$, $C$, $D$ could be determined by
   the normalization conditions,
  %-----------------------------------------------------
   \begin{equation}
  {\int}_{0}^{1}dx\,{\phi}_{\psi}^{v,t}(x) =1,
   \quad
   {\int}_{0}^{1}dx\,{\phi}_{D}^{a,p}(x)=1
   \label{normalization}.
   \end{equation}
  %-----------------------------------------------------

  Here, one may question the validity of the nonrelativistic treatment on
  wave functions of the $D$ mesons, because the motion of the light
  valence quark in the $D$ meson is generally assumed to be relativistic.
  In fact, there are several phenomenological models for $D$ meson
  wave functions, for example, Eq.(30) in Ref. \cite{prd78lv}.
  The $D$ wave function, which is favored by Ref. \cite{prd78lv}
  via fitting with measurements on the $B$ ${\to}$ $DP$ decays
  and often used within the pQCD framework, has the form
  %-----------------------------------------------------
   \begin{equation}
  {\phi}_{D}(x,b) = 6\,x\bar{x}\,\Big\{1+C_{D}(1-2x)\Big\}
  {\exp}\Big\{ -\frac{1}{2}w^{2}b^{2} \Big\}
   \label{wave-dqw},
   \end{equation}
  %-----------------------------------------------------
  where $C_{D}$ $=$ $0.4$ and $w$ $=$ $0.2$ GeV for the $D_{s}$ meson;
  $C_{D}$ $=$ $0.5$ and $w$ $=$ $0.1$ GeV for the $D_{d}$ meson.
  In addition, the same form of Eq.(\ref{wave-dqw}) is widely
  used in many practical calculation without a distinction
  between twist-2 and twist-3 DAs.

  %-----------------------------------------------------
  \begin{figure}[h]
  \includegraphics[width=0.95\textwidth,bb=80 570 525 720]{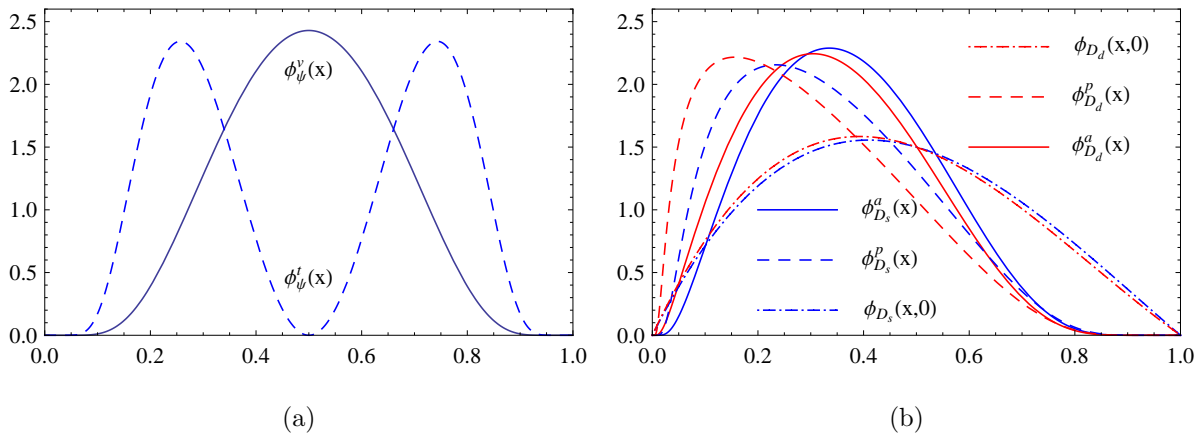}
  \caption{The shape lines of wave functions for the $J/{\psi}$ meson
  in (a) and the $D_{d,s}$ mesons in (b), where the expressions of
  ${\phi}_{\psi}^{v,t}(x)$, ${\phi}_{D}^{a,p}(x)$, and
  ${\phi}_{D}(x,b)$ are given in
  Eqs.(\ref{wave-ccv}---\ref{wave-dqp})
  and Eq.(\ref{wave-dqw}).}
  \label{fig:wave}
  \end{figure}
  %-----------------------------------------------------

  The shape lines of DAs for $J/{\psi}$ and $D_{s,d}$ mesons
  are displayed in Fig. \ref{fig:wave}. It is clearly seen that
  (1) ${\phi}_{\psi}^{v,t}(x)$ for the $J/{\psi}$ meson is symmetric under
  the interchange of momentum fractions $x$ ${\leftrightarrow}$
  $\bar{x}$, and a broad peak of ${\phi}_{D}^{a,p}(x)$ for $D$
  mesons appears at the $x$ $<$ $0.5$ regions, which is basically
  consistent with the scenario that valence quarks in mesons
  might share longitudinal momentum fractions according to their masses.
  (2) Because of the suppression from exponential functions, DAs of
  Eqs.(\ref{wave-ccv}---\ref{wave-dqp})
  fall quickly down to zero at endpoint $x$, $\bar{x}$ ${\to}$ $0$,
  which provides another effective cutoff for soft contributions.
  (3) The flavor symmetry breaking effects between $D_{d}$ and $D_{s}$ mesons,
  and the distinction between twist-2 and twist-3 DAs are apparent in
  Eqs.(\ref{wave-dqa}) and (\ref{wave-dqp}) rather than Eq.(\ref{wave-dqw}).
  Hence, in subsequent calculation, we will take Eqs.(\ref{wave-dqa}) and
  (\ref{wave-dqp}) as the twist-2 and twist-3 DA for the $D$ meson,
  respectively.

  %%%%%%%%%%%%%%%%%%%%%%%%%%
  \subsection{Decay amplitudes}
  \label{sec0205}
\begin{figure}[h]
  \includegraphics[width=0.99\textwidth,bb=75 620 530 725]{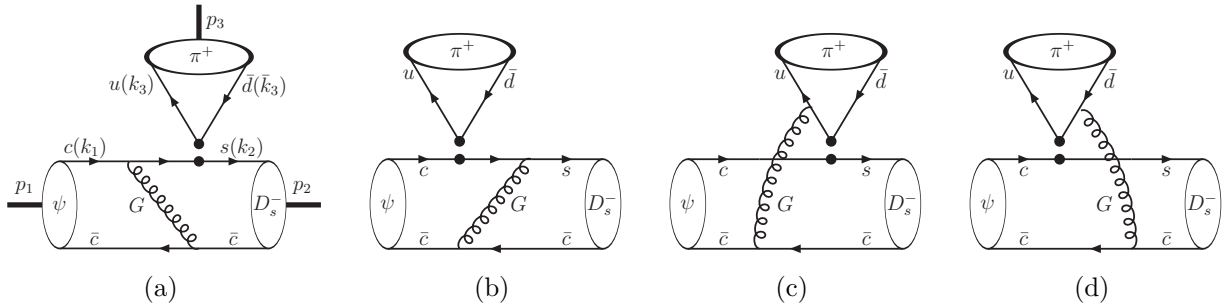}
  \caption{Feynman diagrams for the $J/{\psi}$ ${\to}$ $D_{s}{\pi}$
   decay, where (a) and (b) are factorizable emission diagrams,
   (c) and (d) are nonfactorizable emission diagrams.}
  \label{feynman}
  \end{figure}

  The Feynman diagrams for the $J/{\psi}$ ${\to}$ $D_{s}{\pi}$
  decay within the pQCD framework are shown in Fig.\ref{feynman},
  including factorizable emission topologies (a) and (b)
  where gluon connects $J/{\psi}$ with the $D_{s}$ meson,
  and nonfactorizable emission topologies (c) and (d)
  where gluon couples the spectator quark with
  the emitted pion.

  %-----------------------------------------------------
  %-----------------------------------------------------

  After calculation with the master pQCD formula,
  amplitude for $J/{\psi}$ ${\to}$ $D_{q}P$ decay is written as
  %-----------------------------------------------------
   \begin{equation}
  {\cal A}(J/{\psi}{\to}D_{q}P) =
   \frac{G_{F}}{\sqrt{2}}\, V_{cq_{1}}V_{uq_{2}}^{\ast}\, C_{{\cal A}}
   \sum\limits_{i} {\cal A}_{i}
   \label{eq:amp01},
   \end{equation}
  %-----------------------------------------------------
   \begin{equation}
   C_{{\cal A}} = m_{\psi}\, ({\epsilon}_{\psi}{\cdot}p_{D})\,
   f_{\psi}\,f_{D_{q}}\,f_{P}\,{\pi}\, C_{F}/{N_{c}}
   \label{eq:amp02},
   \end{equation}
  %-----------------------------------------------------
  where the color number $N_{c}$ $=$ $3$ and color factor
  $C_{F}$ $=$ $4/3$; the subscript $i$ on ${\cal A}_{i}$
  corresponds to indices of Fig.\ref{feynman}.
  The expressions of building blocks ${\cal A}_{i}$
  can be found in Appendix \ref{blocks}.

  According to the modeling notation of Ref. \cite{zpc29},
  the longitudinal axial-vector form factor $A_{0}$ for
  $J/{\psi}$ ${\to}$ $D_{q}$ transition is defined as
  \begin{equation}
 {\langle}D(p_{2}){\vert}\,\bar{q}\,{\gamma}_{\mu}{\gamma}_{5}\,c\,
 {\vert}J/{\psi}(p_{1},{\epsilon}_{1}^{\parallel}){\rangle}
 =i\,2\,m_{\psi}\,\frac{{\epsilon}_{1}^{\parallel}{\cdot}q}{q^{2}}\,
    q_{\mu}\, A_{0}(q^{2})
  \label{eq:ffa0},
  \end{equation}
  where the momentum transfer $q$ $=$ $p_{1}$ $-$ $p_{2}$.
  Form factor $A_{0}$ with the pQCD approach can be expressed as
  \begin{equation}
  A_{0}(q^{2})\, =\, -\frac{{\pi}\,C_{F}}{2\,N_{c}} f_{\psi}\,f_{D_{q}}
  \left.\Big\{ {\cal A}_{a}+{\cal A}_{b} \Big\} \right\vert_{a_{1}=1}^{m_{3}^{2}=q^{2}}
  \label{eq:ffa0-pqcd}.
  \end{equation}

  %%%%%%%%%%%%%%%%%%%%%%%%%%%%%%%%%%%%%%%%%%%%%%%%%%%%%%%%%%%
  \section{Numerical results and discussion}
  \label{sec03}

  In the rest frame of the $J/{\psi}$ meson, the branching ratio is defined as
 %-----------------------------------------------------
   \begin{equation}
  {\cal B}r(J/{\psi}{\to}DP) = \frac{1}{12{\pi}}\,
   \frac{p}{m_{\psi}^{2}{\Gamma}_{\psi}}\,
  {\vert}{\cal A}(J/{\psi}{\to}DP){\vert}^{2}
   \label{br}.
   \end{equation}
 %-----------------------------------------------------

  %%%%%%%%%%%%%%%%%%%%%%%%%%%%%%%%%%%%%%%%%%%%%%%%%%%%%%%%%%%
   \begin{table}[h]
   \caption{The numerical values of input parameters.}
   \label{tab:input}
   \begin{ruledtabular}
   \begin{tabular}{c|ll}
   \multirow{2}{*}{CKM parameters\footnotemark[3] \cite{pdg}}
  & $A$          $=$ $0.814^{+0.023}_{-0.024}$,
  & ${\lambda}$  $=$ $0.22537{\pm}0.00061$, \\
  & $\bar{\rho}$ $=$ $0.117{\pm}0.021$,
  & $\bar{\eta}$ $=$ $0.353{\pm}0.013$, \\ \hline
    \multirow{6}{*}{\begin{tabular}{c} mass and \\ decay constants \end{tabular}}
  & $m_{\psi}$  $=$ $3096.916{\pm}0.011$ MeV \cite{pdg},
  & $f_{\psi}$ $=$ $395.1{\pm}5.0$ MeV, \\
  & $m_{D_{s}}$ $=$ $1968.30{\pm}0.11$ MeV \cite{pdg},
  & $f_{D_{s}}$ $=$ $257.5{\pm}4.6$ MeV \cite{pdg}, \\
  & $m_{D_{d}}$ $=$ $1869.61{\pm}0.10$ MeV \cite{pdg},
  & $f_{D_{d}}$ $=$ $204.6{\pm}5.0$ MeV \cite{pdg}, \\
  & $m_{c}$ $=$ $1.67{\pm}0.07$ GeV \cite{pdg},
  & $f_{K}  $ $=$ $156.2{\pm}0.7$ MeV \cite{pdg}, \\
  & $m_{s}$ ${\approx}$ $510$ MeV \cite{uds},
  & $f_{\pi}$ $=$ $130.41{\pm}0.20$ MeV \cite{pdg}, \\
  & $m_{d}$ ${\approx}$ $310$ MeV \cite{uds},
  & ${\Gamma}_{\psi}$ $=$ $92.9{\pm}2.8$ keV \cite{pdg}, \\ \hline
    Gegenbauer moments \cite{npb529}
  & $a_{1}^{K}$ $=$ $-0.06{\pm}0.03$,
  & $a_{2}^{{\pi},K}$ $=$ $0.25{\pm}0.15$.
  \end{tabular}
  \end{ruledtabular}
  \footnotetext[3]{The relations between CKM parameters (${\rho}$, ${\eta}$)
   and ($\bar{\rho}$, $\bar{\eta}$) are \cite{pdg}: $({\rho}+i{\eta})$ $=$
   $\displaystyle \frac{ \sqrt{1-A^{2}{\lambda}^{4}}(\bar{\rho}+i\bar{\eta}) }
  { \sqrt{1-{\lambda}^{2}}[1-A^{2}{\lambda}^{4}(\bar{\rho}+i\bar{\eta})] }$.}
  \end{table}

  %%%%%%%%%%%%%%%%%%%%%%%%%%%%%%%%%%%%%%%%%%%%%%%%%%%%%%%%%%%
   \begin{table}[h]
   \caption{Form factor $A_{0}^{J/{\psi}{\to}D_{q}}$ and
   branching ratios for $J/{\psi}$ ${\to}$ $D_{s}{\pi}$, $D_{s}K$,
   $D_{d}{\pi}$, $D_{d}K$ decays,
   where uncertainties of pQCD results come from scale
   $(1{\pm}0.1)t_{i}$, quark mass $m_{c}$, hadronic parameters
   and CKM parameters, respectively.}
   \label{tab:output}
   \begin{ruledtabular}
   \begin{tabular}{lccccc}
   \multicolumn{1}{c}{Reference} & \cite{ijmpa14}\footnotemark[4] & \cite{ahep2013}
   & \cite{ijmpa30} & \cite{epjc55} & pQCD \\ \hline
    \multicolumn{1}{c}{$A_{0}^{J/{\psi}{\to}D_{s}}(0)$}
  & $0.66$ & $0.71$ & $0.55$ & $0.37$
  & $0.62^{+0.07+0.03+0.01}_{-0.03-0.05-0.01}$ \\ \hline
    \multicolumn{1}{c}{$A_{0}^{J/{\psi}{\to}D_{d}}(0)$}
  & $0.61$ & $0.55$ & $0.50$ & $0.27$
  & $0.53^{+0.06+0.04+0.01}_{-0.02-0.03-0.01}$ \\ \hline
    $10^{10}{\times}{\cal B}r(J/{\psi}{\to}D_{s}{\pi})$
  & $6.1$ & $7.4$ & $4.1$ & $2.0$
  & $4.30^{+0.22+0.36+0.20+0.003}_{-2.08-0.62-0.19-0.003}$ \\ \hline
    $10^{11}{\times}{\cal B}r(J/{\psi}{\to}D_{s}K)$
  & $3.9$ & $5.3$ & $2.3$ & $1.6$
  & $2.69^{+0.12+0.28+0.15+0.014}_{-2.07-0.73-0.15-0.014}$  \\ \hline
    $10^{11}{\times}{\cal B}r(J/{\psi}{\to}D_{d}{\pi})$
  & $3.9$ & $2.9$ & $2.2$ & $0.8$
  & $2.09^{+0.13+0.21+0.13+0.011}_{-1.12-0.30-0.12-0.011}$ \\ \hline
    $10^{12}{\times}{\cal B}r(J/{\psi}{\to}D_{d}K)$
  & ... & $2.3$ & $1.3$ & ...
  & $1.34^{+0.07+0.16+0.09+0.015}_{-1.00-0.17-0.09-0.015}$
  \end{tabular}
  \end{ruledtabular}
  \footnotetext[4]{The updated results are listed in Table 4 of Ref. \cite{ahep2013}.}
  \end{table}

  The numerical values of input parameters are listed in
  Table \ref{tab:input}, where if not specified explicitly,
  their central values will be taken as the default inputs.
  Our numerical results are presented in Table \ref{tab:output},
  where the first uncertainty comes from the choice of the typical
  scale $(1{\pm}0.1)t_{i}$, and the expression of $t_{i}$ is
  given in Eq.(\ref{tab}) and Eq.(\ref{tcd});
  the second uncertainty is from quark mass $m_{c}$;
  the third uncertainty is from hadronic parameters including
  decay constants and Gegenbauer moments; and the fourth
  uncertainty of branching ratio comes from CKM parameters.
  The following are some comments:

  (1)
  The different branching ratios arise mainly from values of form
  factor $A_{0}$ and various theoretical models.
  In Refs. \cite{ijmpa14,ahep2013,ijmpa30},
  the form factor $A_{0}$ is evaluated with the Wirbel-Stech-Bauer
  model \cite{zpc29}. In Ref. \cite{epjc55}, the form
  factor $A_{0}$ is calculated with QCD sum rules.
  The results of Refs. \cite{ijmpa14,ahep2013,epjc55} are based on
  naive factorization approximation. Nonfactorizable effects from
  HME are considered with the QCDF scheme in Ref. \cite{ijmpa30}
  and with the pQCD approach in this paper.
  By and large, branching ratio for a given $J/{\psi}$ ${\to}$
  $D_{s,d}P$ decay has the same order of magnitude with different
  phenomenological models.
  One of the important reasons is that the processes considered here
  are all color-favored, i.e., $a_{1}$-dominated, which is,
  in general, insensitive to nonfactorizable corrections to HME.

  (2)
  There is a clear hierarchical pattern among branching ratios,
  mainly resulting from the hierarchical structure of CKM
  factors in Eqs.(\ref{eq:vcsvud}---\ref{eq:vcdvus}), i.e.,
  \begin{equation}
  {\cal B}r(J/{\psi}{\to}D_{s}{\pi})\, {\gg}\,
  {\cal B}r(J/{\psi}{\to}D_{s}K)\, {\sim}\,
  {\cal B}r(J/{\psi}{\to}D_{d}{\pi})\, {\gg}\,
  {\cal B}r(J/{\psi}{\to}D_{d}K)
  \label{eq:br-4}.
  \end{equation}
  In addition, because of form factors $A_{0}^{J/{\psi}{\to}D_{s}}$
  ${\gtrsim}$ $A_{0}^{J/{\psi}{\to}D_{d}}$ and decay constants
  $f_{K}$ ${\gtrsim}$ $f_{\pi}$, there is generally a relation
  ${\cal B}r(J/{\psi}{\to}D_{s}K)$ ${\gtrsim}$
  ${\cal B}r(J/{\psi}{\to}D_{d}{\pi})$ with different models.
  Above all, the Cabibbo- and color-favored $J/{\psi}$ ${\to}$
  $D_{s}{\pi}$ decay has a relatively large branching ratio among
  nonleptonic $J/{\psi}$ weak decays, about ${\sim}$
  ${\cal O}(10^{-10})$, which might be potentially accessible at
  the future high-luminosity experiments, such as super tau-charm
  factory, LHC and SuperKEKB.

  %-----------------------------------------------------
  \begin{figure}[h]
  \includegraphics[width=0.95\textwidth,bb=80 570 525 720]{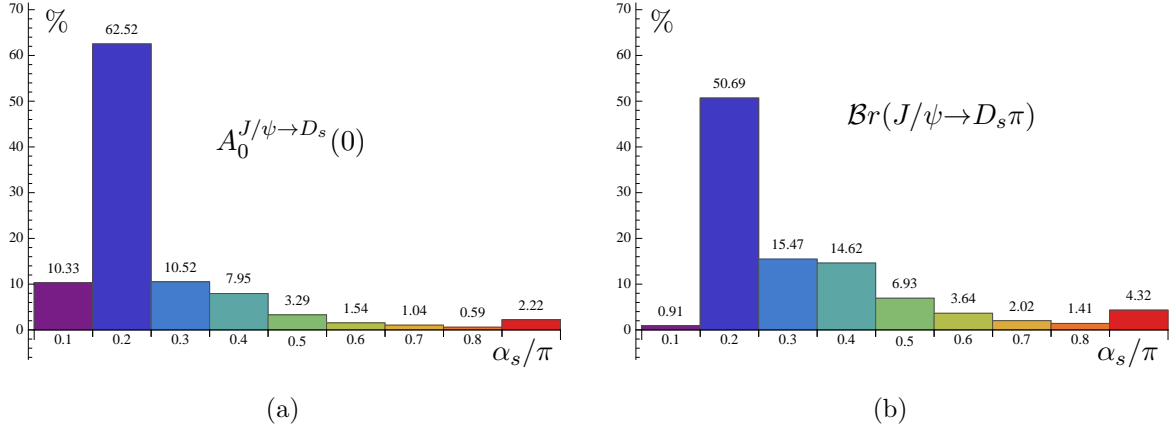}
  \caption{Contributions to form factor $A_{0}^{J/{\psi}{\to}D_{s}}$
  in (a) and branching ratio ${\cal B}r(J/{\psi}{\to}D_{s}{\pi})$ in (b)
  from different regions of ${\alpha}_{s}/{\pi}$ (horizontal axes),
  where the numbers over the histograms denote the percentages of the
  corresponding contributions.}
  \label{fig:as}
  \end{figure}
  %-----------------------------------------------------

  (3)
  It is usually thought that the scale of the $c$ quark mass is not large enough,
  besides the large mass of final states, maybe the momentum transferred in
  the $J/{\psi}$ ${\to}$ $D_{s,d}P$ decay is soft rather than hard.
  One might naturally question the validness of the pQCD approach and the
  reliability of the perturbative calculation.
  Hence, it is very necessary to check what percentage of contributions
  come from the (non)perturbative domain. Taking the $J/{\psi}$ ${\to}$ $D_{s}{\pi}$
  decay as an example, contributions to form factor $A_{0}^{J/{\psi}{\to}D_{s}}$
  and branching ratio ${\cal B}r(J/{\psi}{\to}D_{s}{\pi})$ from different
  ${\alpha}_{s}/{\pi}$ regions are plotted in Fig.\ref{fig:as}.
  It is easily seen that more than 90\% [80\%] of the contributions of
  $A_{0}^{J/{\psi}{\to}D_{s}}$ [${\cal B}r(J/{\psi}{\to}D_{s}{\pi})$]
  come from ${\alpha}_{s}/{\pi}$ ${\le}$ 0.4 regions, which implies that
  the $J/{\psi}$ ${\to}$ $D_{s,d}P$ decays might be computable with the pQCD
  approach. Additionally, as it is well known that
  ${\cal B}r(J/{\psi}{\to}D_{s}{\pi})$ ${\propto}$
  ${\vert}A_{0}^{J/{\psi}{\to}D_{s}}{\vert}^{2}$,
  however, the probability distribution of ${\cal B}r(J/{\psi}{\to}D_{s}{\pi})$
  in Fig.\ref{fig:as}(b) is different from that of $A_{0}^{J/{\psi}{\to}D_{s}}$
  in Fig.\ref{fig:as}(a). In the bin of ${\alpha}_{s}/{\pi}$ ${\in}$ $[0.1,0.2]$,
  the percentage in Fig.\ref{fig:as}(a) is larger than that in Fig.\ref{fig:as}(b),
  while the case is reversed in other bins. One of the critical factors is the
  Wilson coefficients $C_{1,2}$ or $a_{1}$ whose absolute values decrease
  along with the increase of renormalization scale ${\mu}$.
  As it is discussed \cite{pqcd1,pqcd2,pqcd3}, a
  perturbative calculation with the pQCD approach is influenced
  by many factors, for example, the choice of typical scale $t$,
  Sudakov factors, models of wave functions, etc., which deserve much
  attention and further study but are beyond the scope of this paper.

  (4)
  There are many uncertainties on branching ratios, especially
  from scale $t$ and wave functions ($m_{c}$ and hadronic
  parameters). In addition, other factors, such as the final
  state interactions which are usually assumed to be important
  and necessary for $c$ quark decay, different phenomenological
  models for wave functions, and so on, are not
  properly considered here, but deserve massive dedicated
  study. Our results just provide an order of magnitude
  estimation on the branching ratio.

  %%%%%%%%%%%%%%%%%%%%%%%%%%%%%%%%%%%%%%%%%%%%%%%%%%%%%%%%%%%
  \section{Summary}
  \label{sec04}
  The nonleptonic $J/{\psi}$ weak decay is allowable within the
  standard model. In this paper, we investigated the charm-changing
  $J/{\psi}$ ${\to}$ $D_{s,d}{\pi}$, $D_{s,d}K$ weak decays with
  pQCD approach.
  It is found that the estimated branching ratio for the Cabibbo-
  and color-favored $J/{\psi}$ ${\to}$ $D_{s}{\pi}$ decay can reach
  up to ${\cal O}(10^{-10})$, which might be promisingly measurable
  in future experiments.

  %%%%%%%%%%%%%%%%%%%%%%%%%%%%%%%%%%%%%%%%%%%%%%%%%%%%%%%%%%%
  \section*{Acknowledgments}
  We thank Professor Dongsheng Du (IHEP@CAS) and Professor Yadong
  Yang (CCNU) for helpful discussion. We thank the referees for their comments.
  The work is supported by the National Natural Science Foundation
  of China (Grant Nos. 11547014, 11475055, 11275057 and U1332103).

  %%%%%%%%%%%%%%%%%%%%%%%%%%%%%%%%%%%%%%%%%%%%%%%%%%%%%%%%%%%
  \begin{appendix}
  \section{Building blocks of decay amplitudes}
  \label{blocks}
  The explicit expressions of building blocks ${\cal A}_{i}$
  are collected as follows:
  %-----------------------------------------------------
   \begin{eqnarray}
  {\cal A}_{a} &=&
  {\int}_{0}^{1}dx_{1}  {\int}_{0}^{1}dx_{2}
  {\int}_{0}^{\infty}b_{1}db_{1}  {\int}_{0}^{\infty}b_{2}db_{2}\,
  {\phi}_{\psi}^{v}(x_{1})\, E_{ab}(t_{a})\,
  H_{ab}({\alpha},{\beta}_{a},b_{1},b_{2})
   \nonumber \\ &{\times}&
  {\alpha}_{s}(t_{a})\, a_{1}(t_{a})
   \Big\{ {\phi}_{D}^{a}(x_{2})\, \Big[ (t+s)\,\bar{x}_{2}
   -(t+u) \Big] - 2\,m_{2}\,m_{c}\, {\phi}_{D}^{p}(x_{2}) \Big\}
   \label{amp:a},
   \end{eqnarray}
  %-----------------------------------------------------
   \begin{eqnarray}
  {\cal A}_{b} &=&
  {\int}_{0}^{1}dx_{1}  {\int}_{0}^{1}dx_{2}
  {\int}_{0}^{\infty}b_{1}db_{1}  {\int}_{0}^{\infty}b_{2}db_{2}\,
  E_{ab}(t_{b})\, H_{ab}({\alpha},{\beta}_{a},b_{2},b_{1})\,
  {\alpha}_{s}(t_{b})\, a_{1}(t_{b})
   \nonumber \\ &{\times}&
    \Big\{  {\phi}_{\psi}^{v}(x_{1})\, {\phi}_{D}^{a}(x_{2})\,
   \Big[ (t-s)-(t-u)\,\bar{x}_{1} \Big]
 -{\phi}_{\psi}^{t}(x_{1})\, {\phi}_{D}^{p}(x_{2})\,
   4\,m_{1}\,m_{2}\,x_{1} \Big\}
   \label{amp:b},
   \end{eqnarray}
  %-----------------------------------------------------
   \begin{eqnarray}
  {\cal A}_{c} &=& \frac{2}{N_{c}}
  {\int}_{0}^{1}dx_{1} {\int}_{0}^{1}dx_{2} {\int}_{0}^{1}dx_{3}
  {\int}_{0}^{\infty}db_{1}  {\int}_{0}^{\infty}b_{2}db_{2}
  {\int}_{0}^{\infty}b_{3}db_{3}
   \nonumber \\ &{\times}&
   {\phi}_{P}^{a}(x_{3})\, E_{cd}(t_{c})\,
   H_{cd}({\alpha},{\beta}_{c},b_{2},b_{3})\,
   {\alpha}_{s}(t_{c})\,C_{2}(t_{c})
   \nonumber \\ &{\times}&
   \Big\{ {\phi}_{\psi}^{v}(x_{1})\, {\phi}_{D}^{a}(x_{2})\,
   \Big[ t\,(\bar{x}_{1}-\bar{x}_{2})+s\,(\bar{x}_{2}-x_{3}) \Big]
   \nonumber \\ & &+
  {\phi}_{\psi}^{t}(x_{1})\, {\phi}_{D}^{p}(x_{2})\,
   m_{1}\,m_{2}\,(\bar{x}_{2}-\bar{x}_{1}) \Big\}\,
  {\delta}(b_{1}-b_{2})
   \label{amp:c},
   \end{eqnarray}
  %-----------------------------------------------------
   \begin{eqnarray}
  {\cal A}_{d} &=& \frac{2}{N_{c}}
  {\int}_{0}^{1}dx_{1} {\int}_{0}^{1}dx_{2} {\int}_{0}^{1}dx_{3}
  {\int}_{0}^{\infty}db_{1}  {\int}_{0}^{\infty}b_{2}db_{2}
  {\int}_{0}^{\infty}b_{3}db_{3}
   \nonumber \\ &{\times}&
  {\phi}_{P}^{a}(x_{3})\,
  E_{cd}(t_{d})\, H_{cd}({\alpha},{\beta}_{d},b_{2},b_{3})\,
  {\alpha}_{s}(t_{d})\, C_{2}(t_{d})
   \nonumber \\ &{\times}&
   {\delta}(b_{1}-b_{2})\, \Big\{
  {\phi}_{\psi}^{v}(x_{1})\, {\phi}_{D}^{a}(x_{2})\, s\, (x_{2}-x_{3})
   \nonumber \\ & &+
  {\phi}_{\psi}^{t}(x_{1})\, {\phi}_{D}^{p}(x_{2})\,
   m_{1}\,m_{2}\,(x_{1}-x_{2}) \Big\}
   \label{amp:d},
   \end{eqnarray}
  %-----------------------------------------------------
  where $b_{i}$ is the conjugate variable of the transverse
  momentum $k_{iT}$; ${\alpha}_{s}$ is the QCD running coupling;
  $a_{1}$ $=$ $C_{1}$ $+$ $C_{2}/N_{c}$.

  The hard scattering function $H_{i}$ and Sudakov factor $E_{i}$
  are defined as follows.
  %-----------------------------------------------------
   \begin{equation}
   H_{ab}({\alpha},{\beta},b_{i},b_{j}) =
   K_{0}(b_{i}\sqrt{-{\alpha}})
   \Big\{ {\theta}(b_{i}-b_{j})
   K_{0}(b_{i}\sqrt{-{\beta}})
   I_{0}(b_{j}\sqrt{-{\beta}})
   + (b_{i}{\leftrightarrow}b_{j}) \Big\}
   \label{hab},
   \end{equation}
  %-----------------------------------------------------
   \begin{eqnarray}
   H_{cd}({\alpha},{\beta},b_{2},b_{3}) &=&
   \Big\{ {\theta}(-{\beta}) K_{0}(b_{3}\sqrt{-{\beta}})
  +\frac{{\pi}}{2} {\theta}({\beta}) \Big[
   iJ_{0}(b_{3}\sqrt{{\beta}})
   -Y_{0}(b_{3}\sqrt{{\beta}}) \Big] \Big\}
   \nonumber \\ &{\times}&
   \Big\{ {\theta}(b_{2}-b_{3})
   K_{0}(b_{2}\sqrt{-{\alpha}})
   I_{0}(b_{3}\sqrt{-{\alpha}})
   + (b_{2}{\leftrightarrow}b_{3}) \Big\}
   \label{hcd},
   \end{eqnarray}
  %-----------------------------------------------------
  %-----------------------------------------------------
   \begin{equation}
   E_{ab}(t)\ =\ {\exp}\{ -S_{\psi}(t)-S_{D}(t) \}
   \label{sudakov-f},
   \end{equation}
  %-----------------------------------------------------
   \begin{equation}
   E_{cd}(t)\ =\ {\exp}\{ -S_{\psi}(t)-S_{D}(t)-S_{P}(t) \}
   \label{sudakov-n},
   \end{equation}
  %-----------------------------------------------------
   \begin{equation}
   S_{\psi}(t)\ =\
   s(x_{1},p_{1}^{+},1/b_{1})
  +2{\int}_{1/b_{1}}^{t}\frac{d{\mu}}{\mu}{\gamma}_{q}
   \label{sudakov-cc},
   \end{equation}
  %-----------------------------------------------------
   \begin{equation}
   S_{D}(t)\ =\
   s(x_{2},p_{2}^{+},1/b_{2})
  +2{\int}_{1/b_{2}}^{t}\frac{d{\mu}}{\mu}{\gamma}_{q}
   \label{sudakov-cq},
   \end{equation}
  %-----------------------------------------------------
   \begin{equation}
   S_{P}(t)\ =\
   s(x_{3},p_{3}^{+},1/b_{3})
  +s(\bar{x}_{3},p_{3}^{+},1/b_{3})
  +2{\int}_{1/b_{3}}^{t}\frac{d{\mu}}{\mu}{\gamma}_{q}
   \label{sudakov-ds},
   \end{equation}
  %-----------------------------------------------------
  where $J_{0}$ and $Y_{0}$ ($I_{0}$ and $K_{0}$) are the
  (modified) Bessel function of the first and second kind,
  respectively; the expression of $s(x,Q,1/b)$ can be
  found in the appendix of Ref.\cite{pqcd1};
  ${\gamma}_{q}$ $=$ $-{\alpha}_{s}/{\pi}$ is the
  quark anomalous dimension;
  ${\alpha}$ and ${\beta}$ are gluon and quark virtuality,
  respectively, which are listed as follows.
  %-----------------------------------------------------
   \begin{eqnarray}
  {\alpha} &=& \bar{x}_{1}^{2}m_{1}^{2}
            +  \bar{x}_{2}^{2}m_{2}^{2}
            -  \bar{x}_{1}\bar{x}_{2}t
   \label{gluon-q2}, \\
  %-----------------------------------------------------
  {\beta}_{a} &=& m_{1}^{2} - m_{c}^{2}
               +  \bar{x}_{2}^{2}m_{2}^{2}
               -  \bar{x}_{2}t
   \label{beta-fa}, \\
  %-----------------------------------------------------
  {\beta}_{b} &=& m_{2}^{2}
               +  \bar{x}_{1}^{2}m_{1}^{2}
               -  \bar{x}_{1}t
   \label{beta-fb}, \\
  %-----------------------------------------------------
  {\beta}_{c} &=& \bar{x}_{1}^{2}m_{1}^{2}
               +  \bar{x}_{2}^{2}m_{2}^{2}
               +  x_{3}^{2}m_{3}^{2}
   \nonumber \\ &-&
                  \bar{x}_{1}\bar{x}_{2}t
               -  \bar{x}_{1}x_{3}u
               +  \bar{x}_{2}x_{3}s
   \label{beta-fc}, \\
  %-----------------------------------------------------
  {\beta}_{d} &=& x_{1}^{2}m_{1}^{2}
               +  x_{2}^{2}m_{2}^{2}
               +  x_{3}^{2}m_{3}^{2}
    \nonumber \\ &-&
                  x_{1}x_{2}t
               -  x_{1}x_{3}u
               +  x_{2}x_{3}s
   \label{beta-fd}, \\
   t_{a(b)} &=& {\max}(\sqrt{-{\alpha}},\sqrt{-{\beta}_{a(b)}},1/b_{1},1/b_{2})
   \label{tab}, \\
   t_{c(d)} &=& {\max}(\sqrt{-{\alpha}},\sqrt{{\vert}{\beta}_{c(d)}{\vert}},1/b_{2},1/b_{3})
   \label{tcd}.
   \end{eqnarray}
  %-----------------------------------------------------
  \end{appendix}

   %%%%%%%%%%%%%%%%%%%%%%%%%%%%%%%%%%%%%%%%%%%%%%%%%%%%%%%%%%%
  
  \end{document}